\documentclass{aa}

\usepackage[utf8]{inputenc}
\usepackage{color}
\usepackage[usenames,dvipsnames,svgnames,table]{xcolor}
\usepackage{rotating}
\usepackage{txfonts}
\usepackage[]{hyperref}
\hypersetup{colorlinks=true,citecolor=blue}

\definecolor{ora}{RGB}{238,118,0}

\begin{document} 

   \title{The runaway greenhouse radius inflation effect}

  \subtitle{An observational diagnostic to probe water on Earth-size planets and test the Habitable Zone concept}

   \author{Martin Turbet \inst{1} \and David Ehrenreich \inst{1} \and Christophe Lovis \inst{1} \and Emeline Bolmont \inst{1} 
   \and Thomas Fauchez \inst{2,3}}

 \institute{Observatoire astronomique de l’Université de Genève, 51 chemin des Maillettes, 1290 Sauverny, Switzerland
 \email{martin.turbet@unige.ch}
 \and
 NASA Goddard Space Flight Center, Greenbelt, Maryland, USA \and
 Goddard Earth Sciences Technology and Research (GESTAR), Universities Space Research Association, Columbia, Maryland, USA}


 \abstract
   {Planets similar to Earth -- but slightly more irradiated -- are expected to enter into a 
runaway greenhouse state, where all surface water rapidly evaporates, forming an optically 
thick H$_2$O-dominated atmosphere. For Earth, this extreme climate transition is thought to 
occur for a $\sim$~6$\%$ increase only of the solar luminosity, though the exact limit at which the 
transition would occur is still a highly debated topic. In general, the runaway greenhouse is believed to be a fundamental process 
in the evolution of Earth-size, temperate planets. 
Using 1-D radiative-convective climate calculations accounting for thick, hot water vapour-dominated 
atmospheres, we evaluate the transit atmospheric thickness of a post-runaway greenhouse atmosphere, 
and find that it could possibly reach over a thousand kilometers (i.e., a few tens of $\%$ of Earth radius). 
This abrupt radius inflation -- resulting from the runaway-greenhouse-induced transition -- 
could be detected statistically by ongoing and upcoming space missions such as TESS, CHEOPS and PLATO 
(combined with precise radial velocity mass measurements with ground-based spectrographs such as 
ESPRESSO, CARMENES or SPIRou), or even in particular cases in multiplanetary systems 
such as TRAPPIST-1 (when masses and radii will be known with good enough precision).
   This result provides the community with an observational test of 
(1) the concept of runaway greenhouse, that defines the inner edge of the traditional 
Habitable Zone, and the exact limit of the runaway greenhouse transition. In particular, this could provide 
an empirical measurement of the irradiation at which 
Earth analogs transition from a temperate to a runaway greenhouse climate state. This astronomical
 measurement would make it possible to statistically estimate how close Earth is from the runaway greenhouse. 
   (2) the presence (and statistical abundance) of water in temperate, Earth-size exoplanets.}

\maketitle

\titlerunning{The runaway greenhouse radius inflation effect}
\authorrunning{Martin Turbet et al.}

\section{Introduction}

Planets similar to Earth -- but slightly more irradiated -- are expected to experience a runaway greenhouse transition, 
a state in which a net positive feedback between surface temperature, evaporation, and atmospheric opacity 
causes a runaway warming \citep{Ingersoll:1969,Goldblatt:2012}. 
This runaway greenhouse positive feedback ceases only when oceans have completely boiled away, 
forming an optically thick H$_2$O-dominated atmosphere. 
Venus may have experienced a runaway greenhouse in the past \citep{Rasool:1970,Kasting:1984}, and we expect that  
Earth will in around 600~million years as solar luminosity increases by $\sim$~6$\%$ compared to its present-day value \citep{Gough:1981}. 
However, the exact limit at which this 
extreme, rapid climate transition from a temperate climate (with most water condensed on the surface) to 
a post-runaway greenhouse climate (with all water in the atmosphere) 
would occur, and whether or not 
a CO$_2$ atmospheric level increase would affect that limit, 
is still a highly debated topic \citep{Leconte:2013nat, Goldblatt:2013, Ramirez:2014b, Popp:2016}. 
This runaway greenhouse limit is traditionally used to define the inner edge of the Habitable Zone \citep{Kasting:1993,Kopparapu:2013}.

\begin{figure}
    \centering
\includegraphics[width=\linewidth]{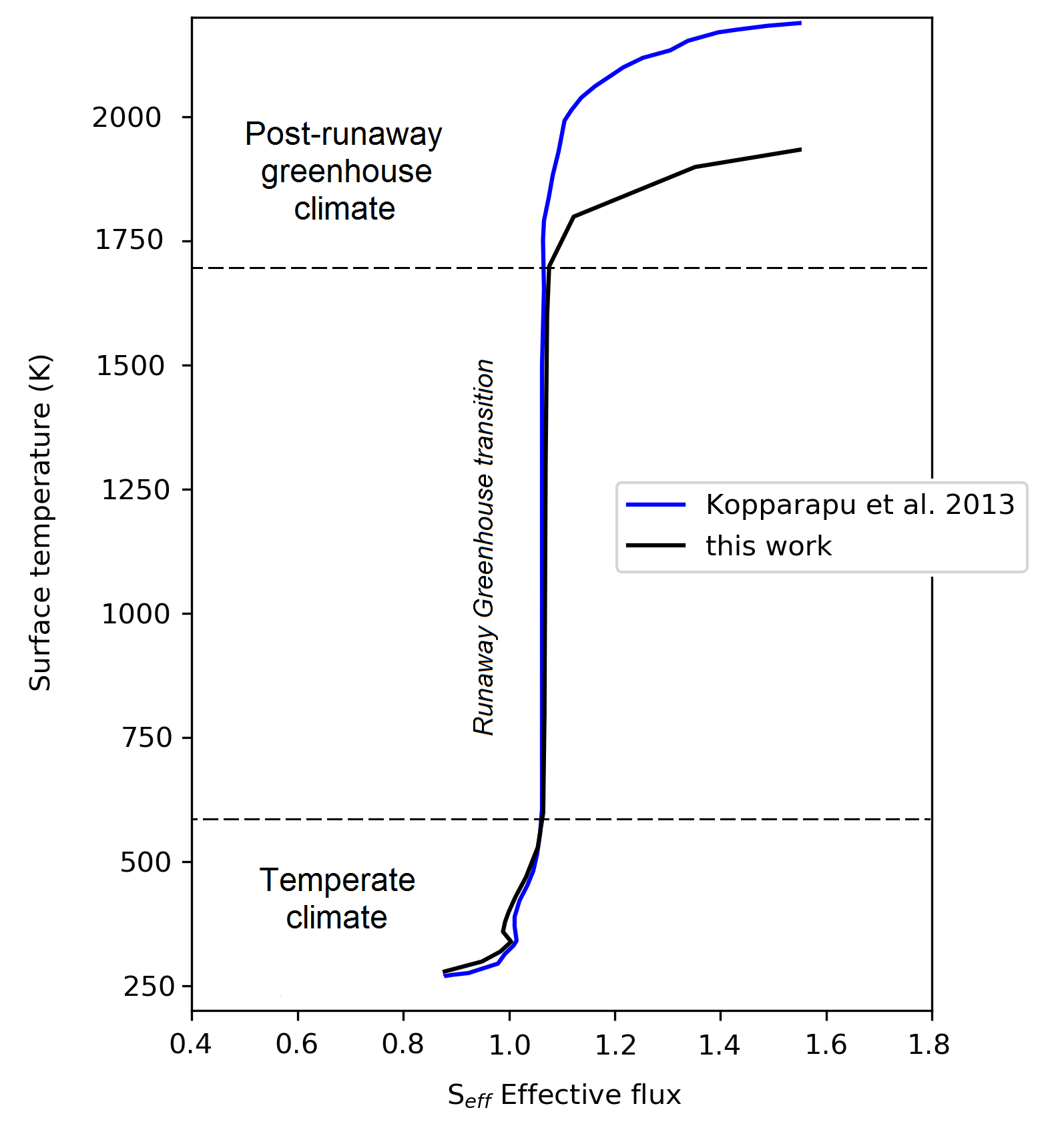}
\caption{Surface temperature of the planet calculated as a function of the effective flux S$_\text{eff}$, 
defined as the stellar flux received by the planet relative to present-day Earth solar flux. 
Here we compare the results of our climate calculations (black line) 
with those presented in the Figure~3c of \citealt{Kopparapu:2013} (blue line).}
\label{Seff_Tsurf_ref}
\end{figure}


Here we take advantage of this runaway greenhouse process to propose a new, innovative observational test of the 
Habitable Zone concept, that could also be used to constrain the presence (and statistical abundance) of water in temperate, Earth-size exoplanets.

\section{The runaway greenhouse radius inflation effect}
\label{runaway_section}

Here we use a 1-D radiative-convective version of the LMD Generic model, that has already been used to simulate the 
runaway greenhouse process on Earth, Mars and exoplanets \citep{Leconte:2013nat,Turbet:2019impact}. 
Our 1D version of the LMD Generic inverse model is a single-column inverse radiative-convective climate model
following the same approach (inverse modeling) as in \citet{Kasting:1984} and more recently as in \citet{Kopparapu:2013}. 
More details on the model can be found in Appendix~\ref{1D_model}.

\begin{figure}
    \centering
\includegraphics[width=\linewidth]{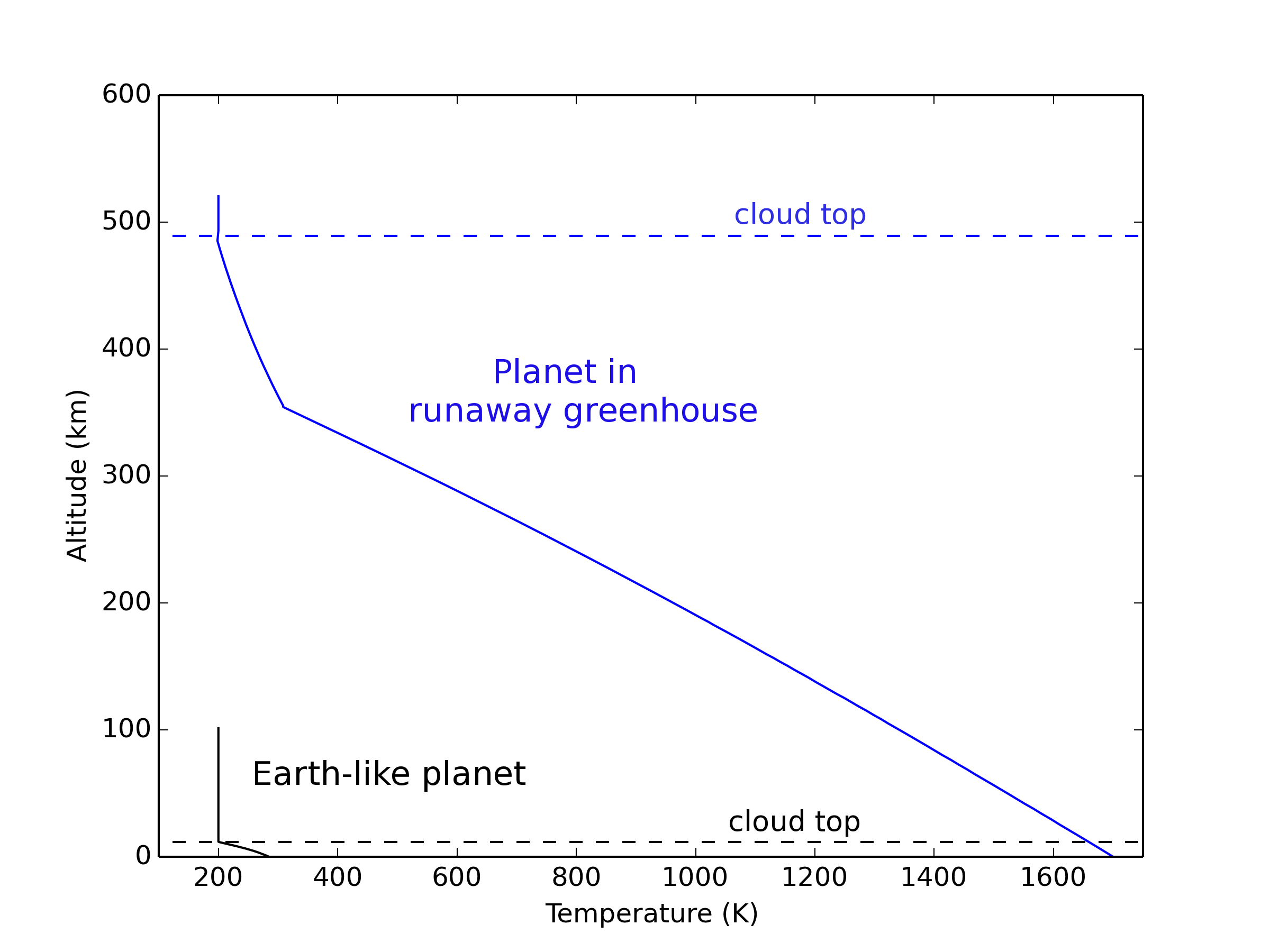}
\caption{Thermal profiles of an Earth-like planet receiving the same insolation than present-day Earth (solid black curve) 
and receiving 6$\%$ more insolation than Earth (solid blue curve). 
In the latter case, the planet is in a post-runaway greenhouse climate state. The top of the cloud layer -- estimated 
using the top of the moist convective layer -- is indicated with dashed lines.}
\label{1D_profiles}
\end{figure}

We use this model to simulate how the climate of an Earth-like planet -- 
covered with a water reservoir equal to that of the Earth ocean water content --
evolves as a function of stellar irradiation (Fig.~\ref{Seff_Tsurf_ref}), following 
the same methodology than used in \citet{Kasting:1993} or more recently in \citet{Kopparapu:2013}. 
We recover the results of \citet{Kopparapu:2013} that an Earth-like 
planet would experience a runaway greenhouse transition
for an insolation equal to $\sim$~1.06~times that of Earth. Moreover,
we calculate that a planet experiencing the runaway greenhouse
would start to equilibrate -- in a post-runaway greenhouse climate state -- for a surface temperature of $\sim$~1700~K,
i.e. only 100~K lower than reported in \citet{Kopparapu:2013}. This difference is likely due to the use of different 
H$_2$O- continuum databases, i.e. BPS \citep{Paynter:2011} versus MT$\_$CKD \citep{Mlawer:2012} here. 

Fig.~\ref{1D_profiles} shows the calculated thermal profiles for an Earth-like planet receiving an insolation equal to 
that of present-day Earth (black line; temperate climate state) and receiving an insolation equal to 1.07 times that of present-day Earth 
(blue line; post-runaway greenhouse climate state). 
Fig.~\ref{1D_profiles} illustrates our main finding that as an Earth-like planet evolves from a temperate to a 
post-runaway greenhouse climate state, the apparent thickness of its 
atmosphere evolves from almost zero up to 500~km. 
Note that the same effect can qualitatively be observed in the Figure~1 of \cite{Goldblatt:2015}, using 
a different 1-D numerical climate model.
The apparent thickness of the planetary atmospheres is defined
here as the top of the cloud layer (assumed to be optically thick),
but we show in Section~\ref{section_water_content} that this result is nearly unchanged
when a cloud-free atmosphere is assumed. 
The transition from a temperate to a post-runaway greenhouse climate would thus produce 
a $\sim$~500~km radius increase. We call this phenomenon the \textit{runaway greenhouse radius inflation effect}.

There are at least four distinct causes that cumulatively produce this radius inflation:
\begin{enumerate}
\item The total mass of the atmosphere increases by a factor of $\sim$~270 (from 1 to $\sim$270~bars, i.e. the estimated 
surface pressure if the Earth ocean water content is entirely vaporized in the atmosphere).
\item The atmosphere becomes much hotter, which increases the atmospheric scale height H~=~$\frac{R T}{\mu g}$ 
(with $R$ the ideal gas constant, $T$ the temperature, $\mu$ the mean molecular mass and $g$ the gravity).
\item The atmosphere is now optically thick at much lower atmospheric pressure, because -- for a given 
atmospheric pressure -- the temperature is much higher, thus increasing the water vapour mixing ratio. This increases in turn 
(i) the water vapour absorption in the upper atmosphere and (ii) the altitude of the top of the cloud layer. 
\item The mean molecular mass $\mu$ decreases (N$_2$ and O$_2$ are heavier than H$_2$O), which increases the atmospheric scale height.
\end{enumerate}

While for Earth the net radius increase is about 500~km, we actually find that the radius increase significantly changes 
depending on the water content. This is explored in the next section.

\section{Dependence on water content}
\label{section_water_content}

The Earth near-surface and surface water reservoir is equal to 1.3$\times$10$^{21}$~kg, 
or 2.7~km GEL (Global Equivalent Layer; i.e. the globally averaged 
depth of the layer that would result from putting all the water at the surface in a liquid phase), or 
270~bars (Equivalent water vapour atmospheric pressure if the entire water reservoir is vaporized in the atmosphere), 
or 0.022$\%$ of the total Earth mass.
But Earth-size planets can potentially have a very different water content than Earth \citep{Raymond:2004,Leger:2004,Tian:2015}. 
Some planets may have started with a low water reservoir 
and may have lost most of it through atmospheric escape (e.g. during the runaway and post-runaway 
greenhouse phase, where water vapour is the dominant gas and 
is thus exposed to photodissociation and subsequent atmospheric escape). 
This is likely what happened to Venus (see the introduction section of \citealt{Way:2016} 
for a recent review). 
But some others may have 
started with a lot of water and retained most of it through their evolution.

\begin{figure}
    \centering
\includegraphics[width=\linewidth]{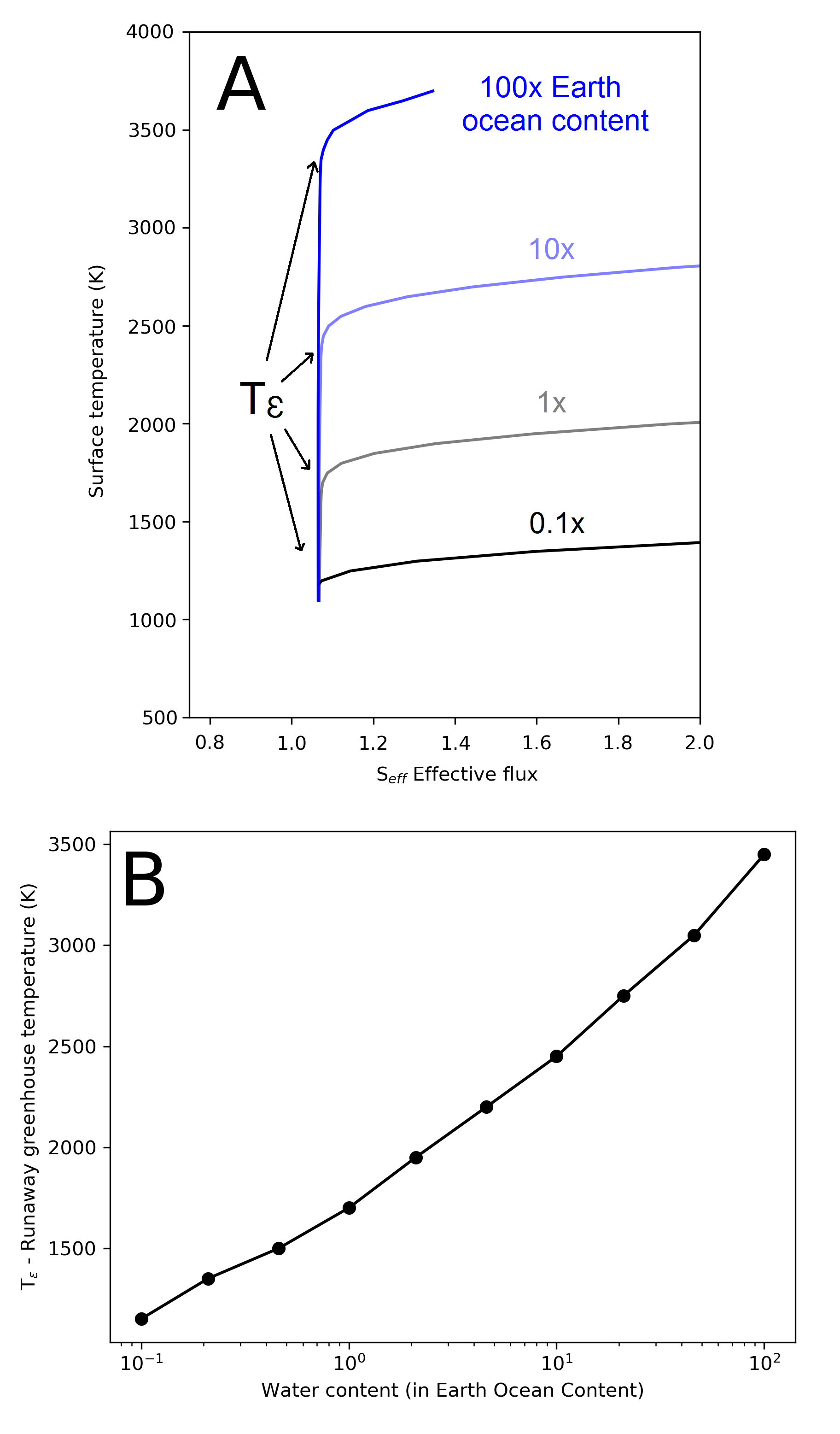}
\caption{Top panel (A): Surface temperature of the planet calculated as a function of the effective flux S$_\text{eff}$ 
(defined as the stellar flux received by the planet relative to present-day Earth solar flux), for 4 distinct total 
water contents (0.1, 1, 10 and 100~times the Earth ocean water content).
Bottom panel (B): Runaway greenhouse temperature (T$_\epsilon$) as a function of the total water content. 
Our best fit is obtained for T$_\epsilon (Wc)$~=~80~$\Big( \log_{10}{Wc} \Big)^2$+700~$\Big( \log_{10}{Wc} \Big)$+1700, with 
$Wc$ the water content in Earth ocean water content units.}
\label{runaway_t_epsilon}
\end{figure}

Using our 1D climate model (see Appendix~\ref{1D_model}), we calculate thermal profiles of Earth-like planets 
in both temperate and post-runaway greenhouse climate states, and for a large range of water contents 
(from 0.1 to 100~times the Earth ocean water content), in order to study how water content changes the amplitude of 
the runaway greenhouse radius inflation effect.
For this, we use the following procedure:
\paragraph{Step~1 -} For a wide range of total water content (from 0.1 to 100$\times$ that of the present-day Earth water ocean content) 
and a wide range of surface temperatures (from 200 to 4000~K), we construct the atmospheric thermal profile 
(temperature and water vapour mixing ratio) using the 1D LMD Generic model described in 
Appendix~\ref{1D_model}.
\paragraph{Step~2 -} For each thermal profile, we perform the radiative transfer (see Appendix~\ref{1D_model}) and compute the 
effective flux S$_\text{eff}$, defined as the stellar flux (relative to the solar flux received by present-day Earth) 
necessary for each simulated planetary atmosphere to be in radiative equilibrium (see Appendix~\ref{1D_model}). 
Note that we did not account for the radius inflation effect in these radiative transfer calculations for steam atmospheres, 
originating from the fact that the average radius for thermal emission can vary significantly from the average radius for 
absorption of solar radiation. However, \cite{Goldblatt:2015} showed that this effect should not affect more than 5$\%$ of 
the flux to which a planet undergoes the runaway greenhouse transition, for a planet the size and mass similar to the Earth.

\paragraph{Step~3 -} We calculate the threshold temperature T$_\epsilon$ \citep{Marcq:2017}, defined here as 
the surface temperature at which the effective flux S$_\text{eff}$ (i.e. the stellar 
flux received by the planet, relative to present-day Earth solar flux, needed for a 
planet to reach radiative equilibrium for a given surface temperature) changes 
from a constant runaway greenhouse regime to a post-runaway regime for 
which the effective flux S$_\text{eff}$ starts to significantly increase 
with surface temperature (see Fig.~\ref{runaway_t_epsilon}A). 
T$_\epsilon$ is used here to define the surface temperature at which a planet that experienced 
a runaway greenhouse transition roughly starts to equilibrate. 

Fig.~\ref{runaway_t_epsilon}B shows how the threshold temperature T$_\epsilon$ -- defining the surface temperature 
of the post-runaway greenhouse atmosphere -- varies as a function of water content. 
We find that the threshold temperature increases significantly with the water content, qualitatively recovering the results of 
\citet{Marcq:2017} and \citet{Ikoma:2018}. The higher the water content, the more opaque the H$_2$O-rich lower atmosphere, 
the higher the surface temperature needs to be to radiate in the visible spectral domain (where H$_2$O is a weak absorber, but 
also a weak emitter), allowing the planet to radiatively equilibrate.


\paragraph{Step~4 -} We calculate the thermal and water vapour mixing ratio 
profiles of the post-runaway greenhouse climate state (defined above), 
as a function first of pressure, and then of altitude by integrating the hydrostatic equation (from the surface to the top of the atmosphere).

\begin{figure}
    \centering
\includegraphics[width=\linewidth]{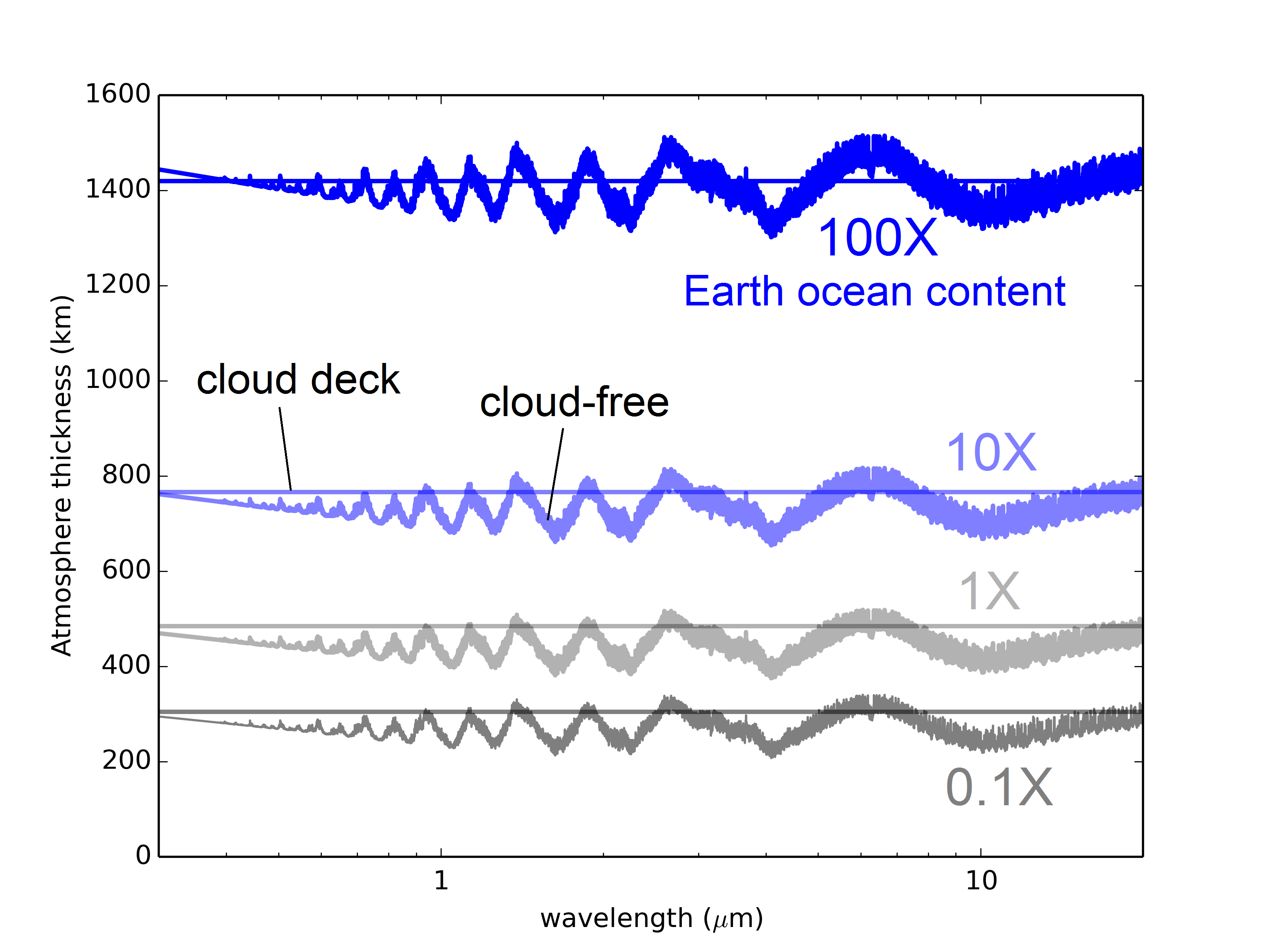}
\caption{Atmospheric thickness of a post-runaway greenhouse planetary atmosphere as a function of wavelength (from 0.3 
to 20~$\mu$m, and for different total water content (from 0.1 to 100$\times$ the Earth ocean content). Atmospheric thickness is plotted 
for cloudy (horizontal lines) and cloud-free scenarios, for comparison.}
\label{spectral_top_atmosphere}
\end{figure}

\paragraph{Step~5 -} We compute the transit atmospheric thickness of the post-runaway greenhouse atmosphere (as seen during a 
transit observation) using two endmember approaches:
\begin{enumerate}
 \item We assume that the transit atmospheric thickness of the atmosphere is defined by the top of the water cloud layer 
(assumed to be optically thick), that we estimate using the top of the moist convective layer. 
This approach likely overestimates the transit atmospheric thickness of the atmosphere.
 \item We compute the high-resolution wavelength-dependent thickness of the atmosphere (assuming a cloud-free atmosphere) using 
the line by line radiative transfer model PUMAS, integrated in the Planetary Spectrum Generator \citep{Villanueva:2018}. 
In this approach, the transit radius is mainly controled by 
H$_2$O band lines absorption and Rayleigh scattering.
This approach likely underestimates the transit atmospheric thickness of the atmosphere.
\end{enumerate} 
Figure~\ref{spectral_top_atmosphere} shows how the 
transit atmospheric thickness of the post-runaway greenhouse atmosphere varies as a function of wavelength for a 
cloud-free versus a cloudy atmosphere. 
Both the cloud-free and cloudy cases give relatively similar transit atmospheric thickness throughout the 
entire spectrum. This is not really surprising, because the upper atmosphere is cold enough (see Fig.~\ref{1D_profiles}) that variations of the 
optically thick pressure level produce weak atmospheric thickness change (if the temperature is low, the atmospheric scale height is also low).

\begin{figure*}
    \centering
\includegraphics[width=\linewidth]{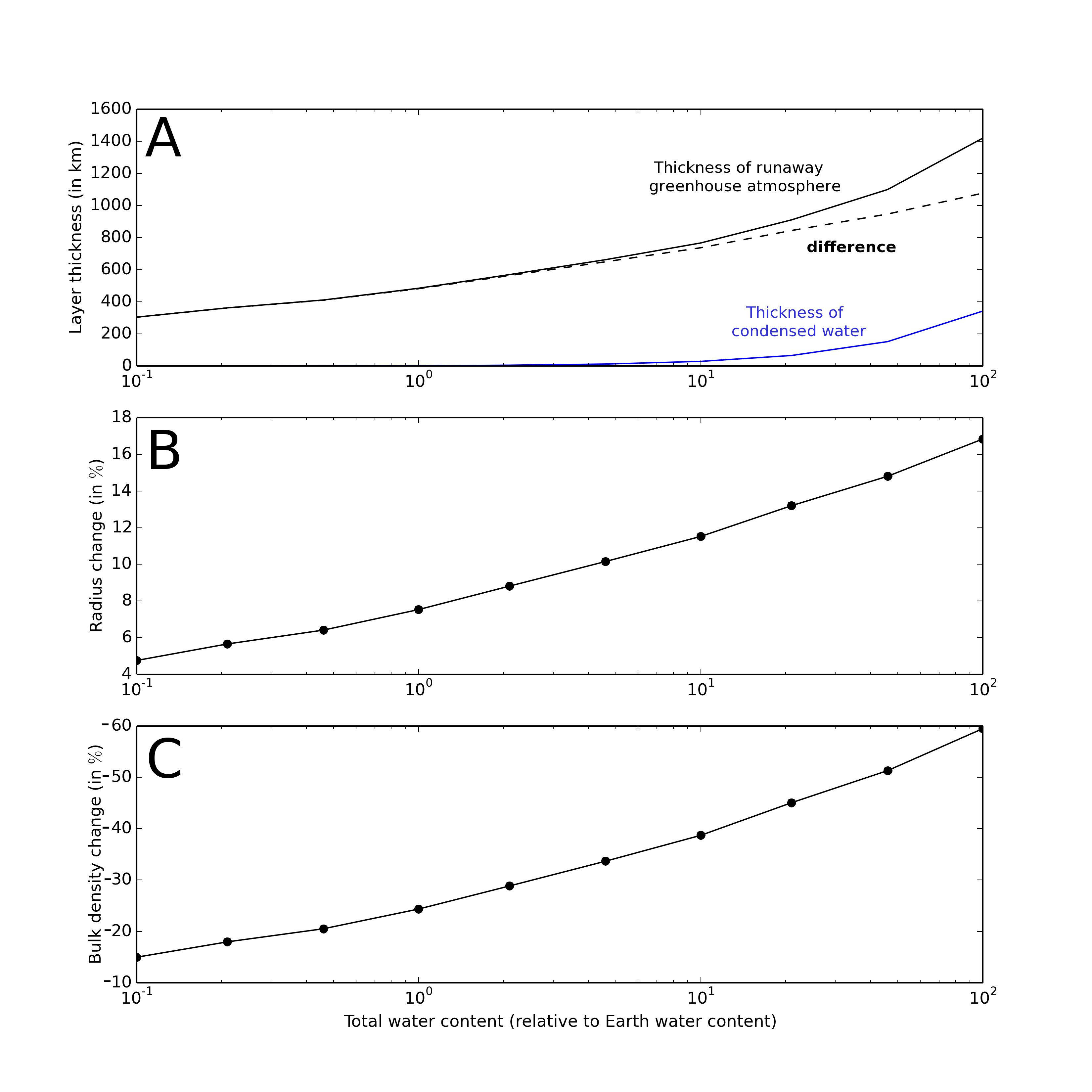}
\caption{Effect of the total water content on: (A) the thickness of the atmosphere (for a planet in post-runaway greenhouse 
climate state, in black) and the thickness of a condensed water layer (for a planet in temperate climate state, in blue). 
The difference has been highlighted in a black, dashed line. (B) the net radius change. (C) the net density change. 
Net radius (B) and density (C) changes are changes relative to Earth radius and density.}
\label{density_change}
\end{figure*}

Figure~\ref{density_change}A shows how the atmospheric thickness of a planet in post-runaway greenhouse state 
$z_\text{atm, post-runaway}$ (solid black line) varies as 
a function of the total water content. When substracting for the thickness of 
the condensed water layer $z_\text{water layer}$ (solid blue line; for present-day Earth, 
$z_\text{water layer}$~=~2.7~km) and the initial atmospheric thickness of a 
planet in temperate climate state $z_\text{atm, pre-runaway}$, this provides 
the net transit radius change $\Delta R$ resulting from the transition between a temperate to a post-runaway greenhouse climate state, calculated 
formally as follows:
\begin{equation}
 \Delta R~=~z_\text{atm, post-runaway}-(z_\text{atm, pre-runaway}+z_\text{water layer})
\end{equation}
For simplicity, and also because the nature of the pre-runaway atmosphere is unknown, 
we assume in our calculations that $z_\text{atm, pre-runaway}$ is equal to 0. 
This seems a reasonable assumption given that it has been shown that the transit 
atmospheric thickness of Earth-like atmospheres should not exceed a few tens of kilometers, 
at least in the visible and infrared spectral domains \citep{Ehrenreich:2006,Kaltenegger:2009,Betremieux:2013}. 
The thickness of the condensed water layer $z_\text{water layer}$ is however accounted for because 
it can have a significant effect on the calculation of the radius change for the 
case of large water content (Fig.~\ref{density_change}A, blue curve).

The net radius change 
varies from $\sim$~300~km (for 0.1$\times$ Earth ocean water content, or 0.002$\%$ of Earth mass) up to 
$\sim$~1100~km (for 100 Earth ocean water content or 2$\%$ of Earth mass), and could possibly 
be even higher for larger water contents, and less massive planets. 
The larger the water content of the planet, the greater the radius inflation.

This net absolute radius change can be translated 
into a net relative radius change (Fig.~\ref{density_change}B) and a net relative density change (Fig.~\ref{density_change}C), 
relative to Earth radius and density values. 
We predict that planets with large water reservoir (at least 2$\%$ of Earth mass or more) would experience -- when transitioning 
from a temperate to a post-runaway greenhouse climate state -- a rapid radius increase 
that could reach or exceed 17$\%$ of Earth radius, and a rapid 
apparent density drop that could reach or exceed 60$\%$ of Earth density. These radius and density changes 
could be further amplified by (i) the runaway-greenhouse-induced 
thermal expansion of the interior \citep{Seager:2007}, 
due to the extreme change of surface temperature boundary condition (from hundreds of Kelvins for a temperate climate planet to 
thousands of Kelvins for a post-runaway greenhouse planet), 
as well as by (ii) the change of internal structure and composition. For example, \cite{Bower:2019} recently 
showed that the melting of the mantle (possibly induced by the runaway greenhouse positive feedback) 
can lead to an additional radius expansion $\sim$~5$\%$.

Moreover, by modifying the internal structure of the planet, the hot post-runaway greenhouse atmosphere could also 
modify gas exchanges between the interior and the atmosphere \citep{Ikoma:2018,Bower:2019}. 
These gas exchanges could affect the amount of volatile (e.g. H$_2$O) outgassing 
and intake from the interior and thus the calculation of the 
post-runaway greenhouse planet radius.
We leave these calculations for a future study. 

The transit net radius change due to the runaway greenhouse transition is orders of magnitude stronger than the 
spectral variation of the atmospheric thickness due to absorption 
by H$_2$O band lines, even in absence of clouds (Fig.~\ref{spectral_top_atmosphere}). 
This is important because it indicates that 
the transit photometry measurement of radius variation due to the runaway-greenhouse-induced transition 
is potentially much easier to do than the transmission spectroscopy measurement of H$_2$O-induced 
spectral changes of the transit radius. The latter have been shown to be very difficult to detect, 
mostly because of clouds \citep{Ehrenreich:2006,Fauchez:2019}. Note however that while transmission spectroscopy 
can be used to identify water in a single planet, here the proposed technique requires to 
perform transit photometry on multiple planets (see next section).

As long as water remains an important component (in condensed form at the surface, or in the form of vapour in the atmosphere) 
of the planet, the transition from a temperate state to a post-runaway greenhouse state can 
in principle occur in both directions, i.e. from temperate to steam atmosphere \citep{Kasting:1993,Kopparapu:2013}, 
but also from steam to temperate atmosphere \citep{Hamano:2013,Lebrun:2013}. 
This indicates that the runaway greenhouse radius inflation (or deflation) 
effect can theoretically occur around any type of host star, 
that the brightness of the host star increases (e.g. for Sun-like stars) or decreases (e.g. for M stars, 
during the extended Pre-Main-Sequence phase) over time.

\begin{figure}
    \centering
\includegraphics[width=\linewidth]{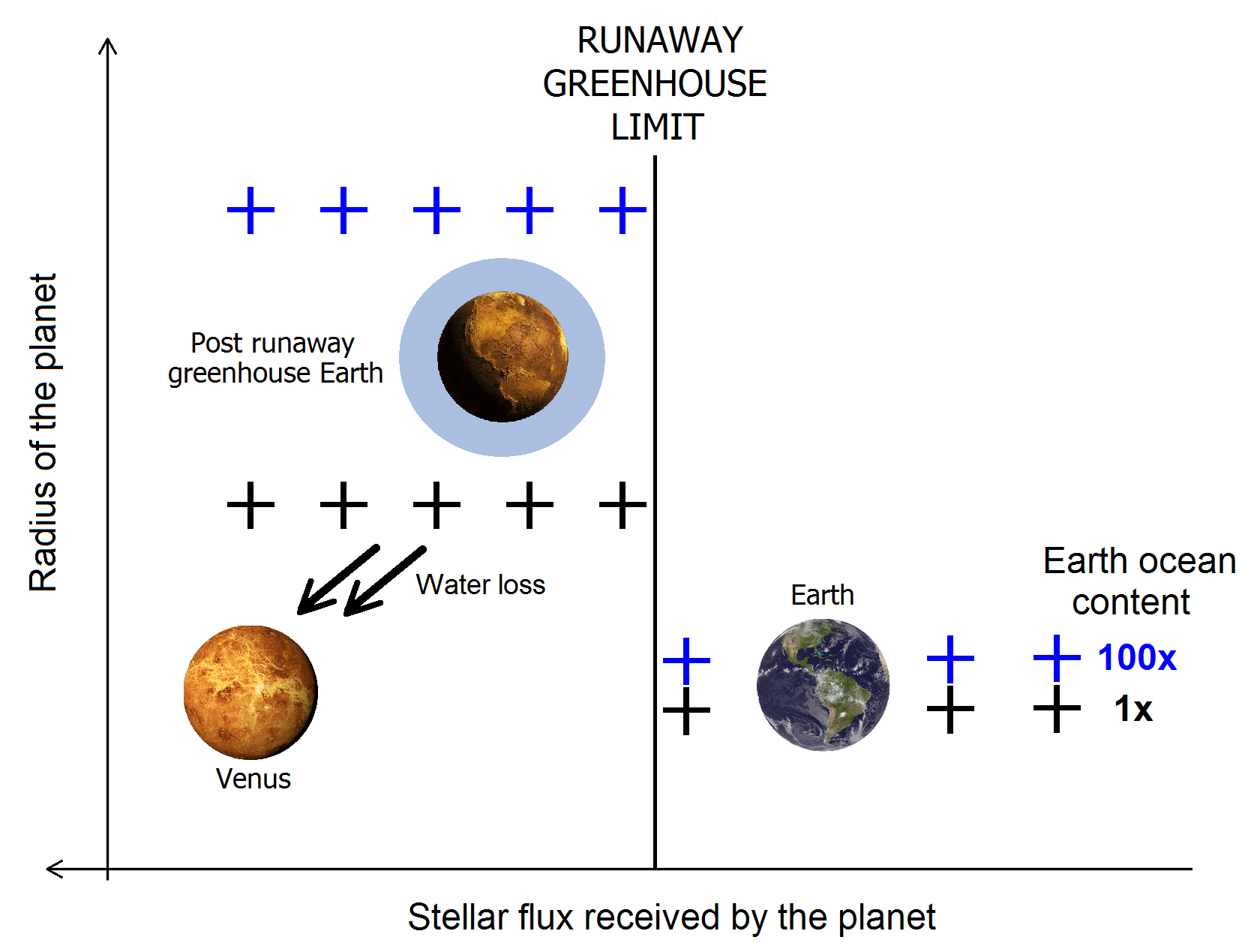}
\caption{Simplified sketch of the runaway greenhouse radius inflation effect, illustrating 
how the transit radius of an Earth-like planet varies as a function of stellar irradiation, assuming two different 
water reservoirs (1 and 100$\times$ the Earth ocean water content). At the runaway greenhouse limit, a planet experiences 
a sharp transition from a temperate to a post-runaway greenhouse climate (or vice versa). This results in a sharp variation of 
the radius and density. While efficient atmospheric escape could erase the runaway greenhouse radius inflation 
effect for planets with limited water reservoir (e.g. Venus), it should have a much less pronounced effect on 
water-rich planets (e.g. with 100x the Earth ocean content), based on theoretical escape rate estimates \citep{Bolmont:2017}.}
\label{sketch_runaway}
\end{figure}

\section{Prospects for an observational test}

This runaway greenhouse radius inflation effect could be tested observationally using at least two different approaches. 

First, upcoming space missions such as TESS \citep{Ricker:2015}, 
CHEOPS \citep{Benz:2018} and PLATO \citep{Rauer:2014} will provide precise measurements of the radii 
of numerous Earth-size exoplanets orbiting on both sides of the theoretical runaway greenhouse limit, 
for M-, K- and potentially even G-type stars.
The radius change $\Delta R / R$ between a pre-- and post--runaway-greenhouse planet (from Fig.~\ref{density_change}B) 
can be converted into a change of the transit depth, $\Delta (\Delta F/F) = ({r_p}^2 (1+\Delta R/R)^2 - {r_p}^2) / {r_\star}^2$, 
where $r_p$ and $r_\star$ are the planetary and stellar radii, respectively \citep{Winn:2010}. Assuming an Earth-size planet (which has a 
transit depth in front of a Sun-like star of $\sim$80~ppm), we calculate $\Delta (\Delta F/F)$ for different stellar radii 
and total water contents of the planet. For G-type stars ($r_\star \approx 1~R_\odot$), the difference in transit depth 
between the pre-- and post--runaway-greenhouse planet is between 8 and 31~ppm depending on the water content of the 
planet (between  0.1 and $100 \times$ the Earth water content, respectively). This values raises to 17--63~ppm for a 
K-type star ($r_\star \approx 0.7~R_\odot$). Red dwarfs offer the largest transit depth differences: an early-type 
M dwarf ($r_\star \approx 0.5~R_\odot$) yields transit depth differences in 32--122~ppm, while a late-type, ultracool 
M dwarf like TRAPPIST-1 ($r_\star \approx 0.1~R_\sun$) offers transit depth differences in 0.08--0.3\% (for 
water contents in  0.1--$100 \times$ the Earth water content, respectively). 
The CHEOPS space mission \citep{Benz:2018} will have a photometric precision of 20~ppm for bright 
stars ($6 \leq V \leq 9$) and 85~ppm for faint stars ($9 < V \leq 12$). It could, in principle, measure 
such transit depth differences pending habitable-zone, Earth-size exoplanets transiting stars bright enough 
are known. This is not the case yet. The TESS mission is expected to yield between 20 and 40 Earth-size planets, 
a handful of which will be found in the habitable zone of M dwarfs (\citealt{Barclay:2018}; see their Fig.~11). 
These objects, if they are similar to TRAPPIST-1, will likely remain beyond the faint limit of CHEOPS in the optical. 
However, one could imagine obtaining very precise infrared radii using space telescopes like ARIEL \citep{Tinetti:2018} or 
the James Webb Space Telescope \citep{Greene:2016}.
The PLATO mission \citep{Rauer:2014} aims at finding many habitable-zone planets for stars as early as the Sun. The 
PLATO sample, for which ages will also be determined, might thus represent a good opportunity to detect and constrain 
the runaway greenhouse radius inflation effect.
These radii measurements will be complemented with follow-up precise radial velocity mass measurements 
using ground-based spectrographs such as ESPRESSO on the VLT \citep{Pepe:2014}, 
CARMENES at Calar Alto Observatory \citep{Quirrenbach:2014} or SPIRou on the CFHT \citep{Artigau:2004}.
Note that the Kepler/K2 sample lacks terrestrial-size, temperate planets around bright stars 
(whose mass can be measured by radial velocity). It cannot therefore be used as is to test 
the runaway greenhouse radius inflation effect.

The abrupt radius inflation -- resulting from the runaway-greenhouse-induced transition -- should produce an extreme planet 
density drop that could be detected statistically using a sample of terrestrial-size planets located on both 
sides of the theoretical inner edge of the Habitable Zone.
The detailed calculation to estimate how many planets are needed could be done following a statistical approach \citep{Bean:2017} 
similar to the detection and characterisation of the gap in the radius distribution due to the transition between 
rocky and volatile-rich exoplanets \citep{Weiss:2014,Rogers:2015,Fulton:2017}.
\citet{Selsis:2007icarus} provides the theoretical framework to evaluate typical sources 
of uncertainties that need to be accounted for in this statistical calculation. 
We acknowledge that the process by which a planet goes into the runaway greenhouse positive feedback 
and inflates its atmosphere has not been studied in time-dependent detail here. Although that time change is 
likely short, it remains an uncertain parameter that must be marginalized over in any statistical inference 
based on the target set for the analysis being proposed here.
Eventually, the true population (most likely a mix of water-rich and dry planets) will not be known in advance, 
and the degree to which the dry planets exist in the sample will dilute the measurements. 
Moreover, the presence of terrestrial-mass planets with a thick H$_2$/He envelope 
\citep{Luger:2015b,Owen:2016} could further dilute the measurements. It is not known if such planets exist 
(i.e. evolved Earth-mass planets with a thick H$_2$/He envelope) and if so, to what extent they are abundant. 
However, these planets are potentially easier to characterize with transmission spectroscopy due to their 
high atmospheric scale height \citep{Dewit:2016,Dewit:2018,Moran:2018}. 
These planets could therefore potentially be identified and removed from the sample. 
In the end, estimating the minimum number of planets needed to begin to observe 
the runaway greenhouse radius inflation effect is a very difficult task, as it depends on many unknown parameters, 
such as the abundance of water in 
the planet sample, as well as the accuracy of density measurements made with the telescopes and ground-based spectrographs 
mentioned above. 
Although the effect could possibly be detected with only a handful of planets if they are all very rich in water 
(and with no H$_2$/He envelope), the possible presence of dry planets and/or planets with a 
H$_2$/He envelope would dilute the expected signal, which could considerably increase 
the number of planets needed to identify the effect, 
possibly to a number far too large to be reached by current and forthcoming telescopes. 
We encourage future studies to evaluate more precisely the minimum number of planets and the accuracy at which 
their density should be known in order to confidently detect the 
runaway greenhouse radius inflation effect, for different water content scenarios.

Secondly, such abrupt density change could be detected locally in multiplanetary systems, 
such as the TRAPPIST-1 resonant chain \citep{Gillon:2016,Gillon:2017,Luger:2017} -
when planet masses will be known with good enough precision. TRAPPIST-1 is a particularly interesting case study, 
because:
\begin{itemize}
 \item it has been speculated that some of the planets may possess large reservoirs of water \citep{Grimm:2018}.
 \item planets are located on both sides of the theoretical runaway greenhouse limit, which is thought to lie between the orbit 
of TRAPPIST-1d and TRAPPIST-1e \citep{Kopparapu:2013,Wolf:2017,Turbet:2018aa}.
\end{itemize}
The detailed analysis of the impact of the runaway greenhouse radius inflation effect within TRAPPIST-1 is currently the subject of a detailed, 
separate study (Turbet et al., publication in preparation).

\section{Discussions}


The detection of such abrupt density change could be used to obtain information on:
\paragraph{1. the concept of runaway greenhouse,} 
that defines the inner edge of the traditional Habitable Zone \citep{Kasting:1993,Kopparapu:2013}. 
The detection of a sharp transition of planet density (locally or statistically) in the 
predicted range of instellation (Fig.~\ref{sketch_runaway}) would indeed validate the concept of runaway greenhouse and Habitable Zone.
In particular, the abrupt density change could be used to estimate 
the limit of the runaway greenhouse transition, possibly as a function of the type of host star. 
 This would make it possible to empirically estimate the position of the inner edge of the Habitable Zone.
 More importantly, this could provide 
an empirical measurement of the irradiation at which Earth analogs orbiting a Sun-like star transition to runaway greenhouse. 
This is especially relevant to the PLATO mission, which aims at detecting Earth-size planets orbiting in the Habitable Zone of 
Sun-like stars. 
This astronomical measurement would make it possible to statistically estimate how close Earth is from the runaway greenhouse.
\paragraph{2. the presence (and statistical abundance) of water in temperate, Earth-size exoplanets,} 
depending on the amplitude of the measured density variation. 
For a given planet sample (e.g. assuming a given range of host star type), 
the runaway greenhouse radius inflation effect can only be detected if water has remained abundant on 
most planets located on both sides of the runaway greenhouse irradiation limit. The ability of planets 
to retain their water depends on (i) the efficiency of initial water accretion during planetary formation 
and on (ii) the efficiency of atmospheric escape processes during planetary evolution \citep{Tian:2015}. 
The latter is particularly important for post-runaway greenhouse atmospheres for which water is dominant 
in the upper atmosphere and is exposed to photolysis and thus to important water loss to space 
\citep{Kasting:1988}. For instance, a steam atmosphere planet (with Earth-like bulk properties) orbiting 
at the location of the runaway greenhouse limit could experience a water loss as high as a few Earth 
ocean water content per billion year \citep{Selsis:2007,Hamano:2015,Bolmont:2017}, 
assuming the energy-limited approximation \citep{Watson:1981}. Note that the runaway greenhouse 
radius inflation could increase the escape rate by a factor of $\sim$~(1.17)$^3$~=~1.6 maximum \citep{Erkaev:2007} 
for an extreme radius increase of 17$\%$ (for 100$\times$ the Earth water ocean content). As a result, planets 
that started with an Earth-like water content or lower can lose all their water rather quickly, i.e. in 
just a few hundred million years.
However, both water atmospheric loss and water delivery during planetary formation are highly unconstrained processes. 
In particular, the initial water content of some temperate, terrestrial-size planets could be as high as hundreds of 
Earth ocean water content \citep{Raymond:2004,Leger:2004,Tian:2015}, making atmospheric 
escape processes inefficient to remove water from 
water-rich (i.e. substantially richer than the present-day Earth ocean water content) 
steam planets receiving irradiation not significantly 
higher than the runaway greenhouse irradiation limit,  based on the aforementioned water loss estimates. 
As a result, planets that started with significantly more water than Earth or that are robust to water 
loss processes should remain water-rich throughout their entire evolution. They are therefore the ideal 
type of planets for detecting and characterizing the runaway greenhouse radius inflation effect.
In general, if the sampled planets have mainly a water-rich composition, then we predict that the runaway 
greenhouse radius inflation should be strong. Conversely, if the sampled planets are mainly dry (e.g. 
because they have lost all their water due to atmospheric escape ; possibly as in the Solar System, with Venus 
for which efficient atmospheric escape has likely swept almost all of the initial water reservoir) then the 
predicted radius inflation should be absent in the sampled population. 
Even a non-detection of the runaway greenhouse radius inflation effect could therefore be used to obtain 
constraints on (i) the maximum water content of the sampled population of temperate Earth-size planets, 
as well as (ii) on the minimal efficiency of atmospheric escape processes (Fig.~\ref{sketch_runaway}).
We acknowledge these two constraints (i) and (ii) are degenerate and a 
detailed analysis would be needed to estimate these parameters and their uncertainties.

\begin{acknowledgements}
This project has received funding from the European Research Council (ERC) under 
the European Union’s Horizon 2020 research and innovation program (grant agreement No. 724427/FOUR ACES). 
This project has received funding from the European Union’s Horizon 2020 research and 
innovation program under the Marie Sklodowska-Curie Grant Agreement No. 832738/ESCAPE.
M.T. thanks Jeremy Leconte, Franck Selsis, Tristan Guillot, Romain Allart and Baptiste Lavie as well as all participants of 
the WHAM meetings for useful discussions related to this work.
M.T. thanks Ravi Kopparapu for his useful feedback on the manuscript.

\end{acknowledgements}

\bibliographystyle{aa} 
\bibliography{biblio} 

\appendix 

\section{The 1D LMD Generic inverse climate model}
\label{1D_model}

Our 1D version of the LMD Generic model is a single-column inverse radiative-convective climate model following 
the same approach ('inverse modeling') as in \citet{Kasting:1984,Kopparapu:2013,Turbet:2019impact}. 
The atmosphere is decomposed into 200 logarithmically-spaced layers that extend from the ground to the top 
of the atmosphere arbitrarily fixed at 0.1~Pascal. 
The atmosphere, assumed here to be composed of 1~bar of N$_2$ and a variable amount of H$_2$O, is divided in 
at most three physical layers constructed as in \citet{Marcq:2012,Marcq:2017,Pluriel:2019}. From the surface to the top, 
the atmosphere is composed of:
\begin{enumerate}
 \item an unsaturated troposphere, where heat transport is dominated by dry convection. 
The mixing ratio H$_2$O/N$_2$ is constant in this layer.
 \item a moist troposphere, saturated in water vapour, 
where heat transport is dominated by moist convection. It is in this layer that clouds are expected to form. 
 \item a purely radiative, isothermal mesosphere with a temperature fixed to 200~K as previously 
done for similar applications \citep{Kasting:1988,Kopparapu:2013,Leconte:2013nat,Marcq:2017} (i.e. a water-rich mesosphere), 
which have shown that even for very hot surface temperatures,
these atmospheres exhibit rather cool temperatures at their top. In this
layer, the mixing ratio H$_2$O/N$_2$ is constant. 
\end{enumerate}
The moist troposphere is present only if the saturation of water vapor is reached at any altitude. 
If this occurs at the surface, then it is the unsaturated troposphere that does not exist.

The dry and moist convective layers are constructed using \citet{Kasting:1988} formulation of the lapse rate, and 
as done in \citet{Marcq:2017}. While N$_2$ is assumed to behave like an ideal gas, the non-ideal behaviour of H$_2$O 
is accounted by using the Fortan NBS/NRC steam tables \citep{Haar:1984}, as in \citet{Marcq:2017}. 
Comparisons with the IAPWS-95 steam tables \citep{Wagner:2002} gave negligible differences on the thermal profiles, 
at least for the range of applications explored in this work.

Once the thermal profile of the atmosphere is constructed, we compute the radiative transfer in 
both visible and thermal infrared spectral domains, and through the 200~atmospheric layers, using the radiative transfer 
of the LMD Generic Model, historically based on the NASA Ames radiation code 
(\url{https://spacescience.arc.nasa.gov/mars-climate-modeling-group/models.html}). The 
radiative transfer calculations are performed on 38 spectral bands in the thermal infrared and 
36 in the visible domain, using the 'correlated-k' approach \citep{Fu:1992}, as in \citet{Leconte:2013nat}. 
16 non-regularly spaced grid points were used for the g-space integration, where g is the cumulative 
distribution function of the absorption for each band. 
Absorption coefficients are exactly the same than those used in \citet{Leconte:2013nat}, and were designed for 
H$_2$O-dominated atmospheres as expected for planets experiencing a runaway greenhouse transition.
For the radiative transfer calculations, the Sun (assumed to be the host star) is assumed to remain fixed, at a zenith angle of 60$^{\circ}$. 
The surface albedo is arbitrarily fixed to 0.2.

From the radiative transfer calculations, we can derive (1) the Outgoing Longwave Radiation (OLR) and 
(2) the planetary albedo $A_p$. The effective flux S$_\text{eff}$ can then be computed as $\frac{OLR}{(1-A_p)~F_{\oplus}}$, 
with $F_{\oplus}$ the average insolation at the top of the atmosphere on Earth, 
equal to one fourth of the solar constant, i.e. 340.5~W~m$^{-2}$.
The effective flux can be interpreted as the stellar flux (relative to the solar flux received by Earth today) necessary 
for a planet to be in radiative equilibrium, given the OLR of the planet is known.

\end{document}